# Improved PLL Design for Transient Stability Enhancement of Grid Following Converters Based on Lyapunov Method

Fangyuan Sun[1], *Student Member*, *IEEE*, Ruisheng Diao[1*], *Senior Member*, *IEEE*, Ruiyuan Zeng[1], *Student Member*, *IEEE*, Junjie Li[2], Wangqianyun Tang[2]

*Abstract*—Fluctuations in phase angle and frequency under large disturbances can lead to loss of synchronism (LOS) in grid-following (GFL) converters. The power angle and frequency of synchronous generators (SGs) correspond to rotor position and speed, whereas those of converters lack a direct physical counterpart in the real world and can thus be directly adjusted by control methods to prevent loss of synchronization. In this paper, an improved phase-locked loop (PLL) design with reset control for GFL converters is proposed to enhance transient stability. The stability domain (SD) of a GFL converter is first analyzed, and three forms of SD are identified under different short circuit ratios. Secondly, based on the characteristics of the three SD forms, two PLL-reset methods are proposed, including $\omega$ reset and $\omega \& \delta$ reset. Thirdly, to provide the triggering conditions for the PLL-reset control, the Lyapunov function of the GFL converter is constructed based on three methods: the approximation-based Lyapunov method, the Zubov method, and the analytical trajectory reversing method. All these methods are immune to the negative damping problem of PLL dynamics, which makes traditional energy-perspective Lyapunov functions invalid. Finally, the estimation accuracy of the three Lyapunov-based methods is analyzed, and the effectiveness of the PLL-reset control is verified in single-machine and multi-machine case studies.

## I. INTRODUCTION

THE high penetration of renewable energy has become a prominent characteristic of modern power systems, and the grid following (GFL) converters serve as the main grid interface for integrating renewable energy generators. Typically, synchronization of GFL converters is achieved using phase-locked loops (PLLs), which exhibit significantly different transient dynamics compared to synchronous generators (SGs). The loss of synchronism (LOS) caused by PLL dynamics has been reported both in research papers [1] and real power system operations [2].

It has been widely reported that the PLL-dominated dynamic equations of the GFL converter are similar to the swing equation of SGs, and both consist of a mechanical power term, an electrical power term, and a damping term [3]. Unlike the constant damping coefficient of SGs, the damping coefficient of the GFL converter varies with the phase angle and can be negative when the phase angle is large [1]. The Equal Area Criterion (EAC) and the Lyapunov method are the most widely used direct methods of transient stability assessment. Specifically, the traditional Lyapunov function is established from the energy conservation perspective and is usually known as the energy function. When the damping coefficient of GFL is negative, radical estimation of SD or critical clearance time (CCT) may occur, which is unacceptable in transient stability analysis. Two primary methods have been employed to address the negative damping-related issue. However, both methods exhibit certain limitations in terms of accuracy or practicality. 1) Implementation of the transient analysis only in the positive damping domain [4], [5]. However, the results show that the positive damping domain is generally much smaller than real SDs, leading to large conservative errors. In [6], a new Lyapunov function for type-3 wind turbines is proposed based on singular perturbation modeling, and the region that satisfies the Lyapunov condition is analyzed. However, only a permanent fault is considered. 2) Inclusion of the damping term in the Lyapunov function [7], [8]. However, the accurate value of the damping energy consumption or accumulation is highly related to the transient trajectory, which is difficult to calculate. In [7], an iterative EAC method is used to calculate the damping energy, but only the desynchronization during the first swing is considered. At the beginning of the second swing, negative damping may accumulate energy, leading to a loss of synchronization. Additionally, the EAC-based method is only suitable for second-order systems. In [8], the dissipation of the system in the permanent scenario is proved, which makes it feasible to neglect the damping term, but only a permanent fault is considered. Other transient stability methods have been reported for GFL converters [9], such as the perturbation method [10] or the data-driven method [11]. However, they are not easily embedded in control logic due to the complexity of the judgment processes or the construction of Lyapunov functions.

In fact, path-dependent terms in the Lyapunov function also exist in more detailed models, such as the salient pole synchronous generator model in [12] or the system model considering line resistance in [13]. Comprehensive research has been conducted to analyze these systems. In [13], approximation methods are employed to calculate trajectory-dependent terms, such as the ray approximation or trapezoid approximation. New methods of Lyapunov function construction, such as the Zubov method [12] or analytical trajectory reversing method (ATRM) [14], are also proposed. These methods have the potential to be applied to transient stability analysis for GFL converters.

This work was supported by the National Natural Science Foundation of China under Grant U22B6007.

[1]Fangyuan Sun, Ruisheng Diao, and Ruiyuan Zeng are with the ZJU-UIUC Institute, Zhejiang University, Haining, 314400 China (e-mail: fysun_zju@zju.edu.cn, ruiyuan.23@intl.zju.edu.cn).

[2]Junjie Li, and Wangqianyun Tang are with China Southern Grid Electric Power Research Institute Co., Ltd, Guangzhou, 510623 China.

Corresponding author: Prof. Ruisheng Diao, email: ruishengdiao@intl.zju.edu.cn

To avoid LOS of GFL converters, studies have been reported by investigating new control strategies [15]-[20]. In [15] and [16], two voltage normalization methods are adopted in the PLL to decouple the PLL dynamics from variations in the grid voltage amplitude. In [17], a novel GFL control method combining additional frequency control and PLL is proposed, allowing the GFL converter to maintain synchronization under strong, weak, or islanded grid conditions. In [18], an opposite disturbance component is introduced into the traditional GFL control to counteract the negative influence of the PLL. In [19], a novel bandwidth-limiting PLL and parameter tuning method are proposed, which can limit the PLL bandwidth in a desired interval and achieve fast and accurate synchronization with the voltage of the coupling bus. In [1], the influence of the PLL parameters on the stability domain (SD) is analyzed, and a recommendation for the PI controller setting in the PLL is provided. In [20], PLL is reduced to a first-order system by freezing the integrator in the PLL when a fault occurs, and the rate of change of frequency (RoCoF) is used to detect the occurrence of the fault. In [21], an optimized design for demagnetization control is proposed to enhance the transient stability of DFIG-based wind turbines, considering the influence of low-voltage ride-through (LVRT) control.

Based on the above summary, current research efforts primarily focus on developing new structures for PLL or GFL control. The effectiveness of these methods in enhancing transient stability has been proven. However, the following problems still require further investigation: 1) The changed control loops may cause other types of stability issues under special circumstances, such as oscillations with other devices or repeated control switching. 2) The transient stability is enhanced in current methods; however, the avoidance of LOS is not guaranteed in most of these methods. Instability may still occur under severe or specific conditions.

One advantage of converters is that the phase angle is a virtual quantity, which can be arbitrarily changed by the control signal [22]. Therefore, unlike designing a new PLL or GFL structure, there exists another stability enhancement method that actively resets the power angle and frequency to a stable value when the converter state exceeds its stability domain. The advantages of this reset-based stability enhancement method lie in: 1) The PLL with reset mechanism is equivalent to traditional PLL unless the reset is activated, which reduces the risk of other unexpected instability problems; and 2) The reset mechanism is embedded with the transient stability judgment, which gives a theoretical guarantee of the transient stability of the GFL converter.

In this paper, based on the accurate SD assessment, a novel PLL-reset control is proposed to avoid the LOS of the converter following large disturbances. To embed a criterion in the control logic to judge transient stability, Lyapunov functions for the GFL converter are established, which are immune to the aforementioned negative damping problem. The main contributions are summarized as follows:

1) The SD of the GFL converter-connected system is obtained by the trajectory reversing method (TRM), and the SD characteristics under different short-circuit ratios (SCRs) are analyzed.

2) Based on the characteristics of the SDs, two PLL-reset methods are proposed. When the state of the GFL converter exceeds the SD, the frequency or phase angle is reset into the SD to avoid LOS.

3) Three transient stability analysis methods, based on the approximation-based Lyapunov method (ABLM), Zubov method, and ATRM, are first proposed in this paper, providing a trigger condition for the PLL-reset control. All these methods can avoid the negative damping problem caused by the PLL dynamics that make the traditional EAC and Lyapunov methods difficult to apply.

The remainder of the paper is organized as follows. Section II introduces the model of the GFL converter-connected system. In Section III, the SD of the GFL converter is analyzed, and two PLL-reset methods are proposed. In Section IV, three transient analysis methods are introduced. Transient analysis results and the effectiveness of the PLL-reset control methods are tested in Section V. Finally, Section VI draws the conclusion.

## II. SYSTEM MODELING

Fig. 1 shows the single-line diagram of a GFL converter-connected system, and the following assumptions are made. 1) The power grid is represented as an ideal voltage source, whose voltage is given by $U_g \angle 0$. 2) The GFL converter is regarded as a controlled current source with a fixed current amplitude, $I_c$, and the phase angle $\delta_c$ is determined by the phase-locked loop (PLL). The phase angle between $I_c$ and the d-axis of PLL is also fixed and represented by $\varphi_I$. 3) The load is a constant impedance $Z_l$. A fault resistance $R_f$ is connected in parallel to the load when a fault occurs. 4) The influence of frequency fluctuation is small in large power systems, so its influence on reactance is ignored; that is, the reactances in Fig. 1 are constant [8],[23].

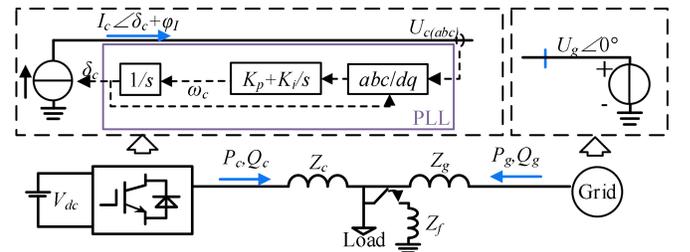

Fig. 1. The diagram of the GFL IBR grid-connecting system.

Synchronization of PLL is realized by detecting the q-axis voltage at the point of common coupling (PCC). From the Kirchhoff's Laws, the PCC voltage of the GFL converter can be calculated as:

$$\left.\begin{array}{l} \dot{U}_l = \dot{U}_g - \dot{I}_g \dot{Z}_g \\ \dot{U}_l = (\dot{I}_g + \dot{I}_c)\dot{Z}_l \\ \dot{U}_c = \dot{U}_l + \dot{Z}_c \dot{I}_c \end{array}\right\} \Rightarrow \dot{U}_c = \dot{Z}_{eq1}\dot{U}_g + \dot{Z}_{eq2}\dot{I}_c \quad (1)$$

$$Z_{eq1}\angle\theta_1 = \frac{\dot{Z}_l'}{\dot{Z}_g + \dot{Z}_l'}, \quad Z_{eq2}\angle\theta_2 = \frac{\dot{Z}_l'\dot{Z}_g}{\dot{Z}_l' + \dot{Z}_g} + \dot{Z}_c$$

Where, $U_c$ is the PCC voltage. $U_l$ is the voltage at the load node. $I_g$ is the output current of the grid. $Z_g$ and $Z_c$ are the grid and converter side impedance. $Z'_l$ is the virtual load impedance. In pre-fault and post-fault periods, $Z'_l = Z_l$, and in during fault period, $Z'_l = Z_l // R_f$. $Z_{eq1}$ and $Z_{eq2}$ are two parameters to simplify the equation, and $\theta_1$ and $\theta_2$ are the phase angles. The upper dot mark indicates that the variable is a vector.

Take the phase of PLL as the reference, the q-axis PCC voltage $U_{cq}$ can be calculated as:

$$U_{cq} = Z_{eq1} U_g \sin(\theta_1 - \delta_c) + Z_{eq2} I_c \sin(\theta_2 + \varphi_I) \quad (2)$$

Where, $\delta_c$ is the output angle of PLL, and $\varphi_I$ is the phase angle between $I_c$ and the d-axis of PLL.

From (2) and the PLL diagram in Fig. 1, the PLL-dominated GFL converter dynamics can be expressed by:

$$F = \begin{cases} \dfrac{d\delta_c}{dt} = \omega_c \\ \dfrac{d\omega_c}{dt} = P_{m,c} - P_{e,c} \sin(\delta_c - \theta_1) - D'_c \omega_c \end{cases} \quad (3)$$

$$P_{m,c} = K_i Z_{eq2} I_c \sin(\theta_2 + \varphi_I)$$
$$P_{e,c} = K_i Z_{eq1} U_g$$
$$D'_c = D_c \cos(\delta_c - \theta_1) = K_p Z_{eq1} U_g \cos(\delta_c - \theta_1)$$

Where, $P_{m,c}$, $P_{e,c}$, and $D'_c$ are the virtual mechanical power, electrical power, and damping coefficient of the dynamic equations. $K_i$ and $K_p$ are the integral and proportional coefficients of PLL. $\omega_c$ is the angular velocity difference between the PLL phase and the grid phase. It is worth noting that when the fault occurs or is cleared, there is a mutation of $U_{cq}$, resulting in a mutation of $\omega_c$ through the proportion part of PLL. This cannot be reflected in (3), and more details can be found in [7],[8].

III. ASSESSMENT OF STABILITY DOMAIN AND PLL-RESET CONTROL

A. Stability Domain Analysis

The SD is represented by $A$, and the boundary is represented by $\partial A$. Obviously, the equilibrium points of the system (3) are [arcsin($P_{m,c}/P_{e,c}$), 0], and those whose first dimension is in the range [$2k\pi$, $2k\pi+\pi/2$] ($k \in \mathbb{Z}$) are stable equilibrium points (SEPs), and those whose first dimension is in the range [$\pi/2+2k\pi$, $2k\pi+\pi$] ($k \in \mathbb{Z}$) are unstable equilibrium points (UEPs). For a dynamic system like (3), its SD can be precisely drawn by the TRM. More details can be found in [23].

The TRM is based on a proven theorem that $\partial A$ is formed by the stable manifolds of UEPs on $\partial A$. For each stable equilibrium point (SEP) of the system (3), there is only one UEP; how to find the stable manifolds of the UEP is the key to obtaining $\partial A$. The TRM is one way to obtain the stable manifolds of UEPs by backward integration, and the procedure is shown as follows:

**Step 1**: Create the system equation according to (3):

$$F' = \begin{cases} \dfrac{d\delta_c}{dt} = -\omega_c \\ \dfrac{d\omega_c}{dt} = -P_{m,c} + P_{e,c} \sin(\delta_c - \theta_1) + D'_c \omega_c \end{cases} \quad (4)$$

The system equation $F'$ in (4) has the same UEP as system $F$ in (3). For arbitrary UEP $x_{ue}=(\delta_{c,ue}, \omega_{c,ue})$ of $F$ and $F'$, its stable manifolds in system $F$ have the same trajectories as the unstable manifolds in the system $F'$, but the directions are reversed.

**Step 2**: Implement the numerical integration of the system $F'$ from the UEP, and obtain the unstable manifolds $W'_u(x_{ue})$ as the boundary $\partial A$.

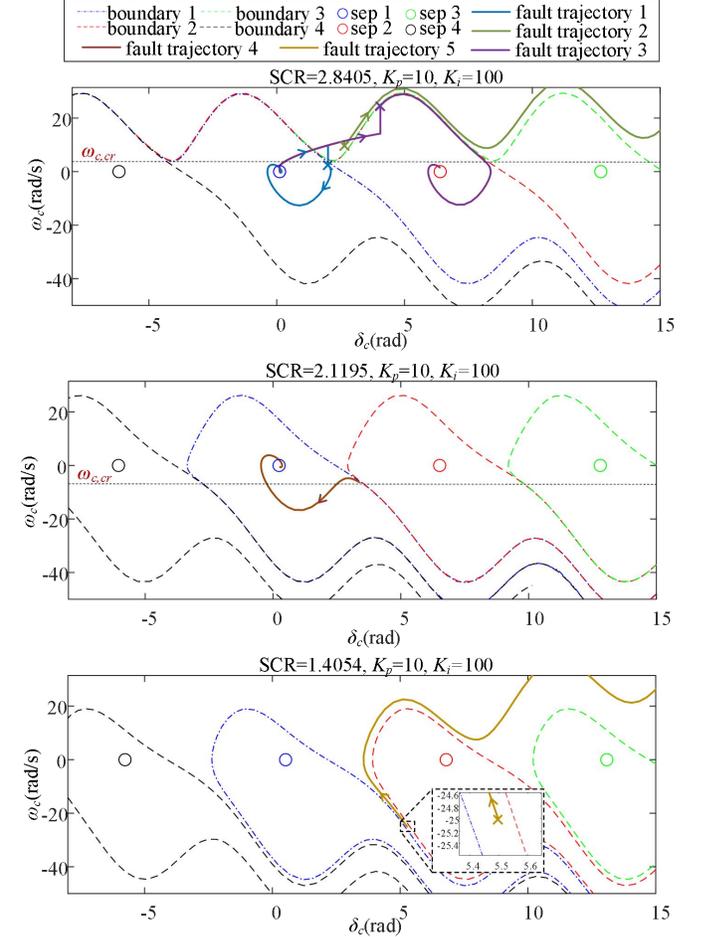

Fig. 2. SDs of the GFL converter under different SCR.

The SDs of the GFL converter under different SCRs are shown in Fig. 2. More parameters are given in the Appendix. The SCR under the parameters given in the Appendix is 2.1195. By changing the parameter $Z_c$ (i.e. $R_c+jX_c$) to 0+j0.24 p.u. and 0+j0.6 p.u., the corresponding SCRs are 2.8405 and 1.4054, respectively. From Fig. 2, three types of SD are observed.

1) When SCR is relatively high (2.8405), the SDs of the SEPs overlap each other, and there is a specific value $\omega_{c,cr}$ that as long as $\omega_c < \omega_{c,cr}$, stability is guaranteed regardless of $\delta_c$.

2) When SCR reduces to 2.1195, the specific value $\omega_{c,cr}$ still exists, but the overlaps between SDs of different SEPs are only part of the boundary. Trajectory 4 (dark brown curve with arrow) is given in the second figure, originating from the initial point (3.6088, -2π), where $\delta_c$ equals the SEP plus π, and $\omega_c$ is less than $\omega_{c,cr}$. Trajectory 4 demonstrates that even with such a large power angle deviation, stability can still be maintained as long as $\omega_c < \omega_{c,cr}$.

3) When SCR further reduces to 1.4054, the SDs of different SEPs do not overlap, and $\omega_{c,cr}$ does not exist. Trajectory 5 in the third figure originates from (5.5, -25). Despite the low $\omega_c$, the initial point is not in the stability region of any SEP, and the trajectory is unstable.

Three fault trajectories are also shown in the first figure. The fault clearing time $t_c$ of the trajectories are 0.32s, 0.38s, and 0.49s. The crossing mark is the converter state ($\delta_c$, $\omega_c$) after $t_c$, and the mutation of $\omega_c$ mentioned in Section II can also be observed. With the increase of $t_c$, the converter state may move from the SD of the initial SEP to the unstable region and then to the SD of the adjacent SEP.

The influence of the control parameters on SD is also analyzed. Firstly, the SDs of GFL converters under different PLL parameters, $K_i$ and $K_p$, are shown in Fig. 3. The SCR is 2.1195. From the first figure, with the increase of $K_i$, the SD shrinks in the $\delta_c$ dimension and expands in the $\omega_c$ dimension. This makes the converter more resistant to frequency disturbance, but the specific value $\omega_{c,cr}$ may not exist. With the increase of $K_p$, the SD expands both in $\delta_c$ dimension and $\omega_c$ dimension. However, under the PLL acceleration scenario, which is a common LOS pattern causing both $\delta_c$ and $\omega_c$ to increase, the SD boundary is interior under larger $K_p$, which is highlighted by the dashed red circle in Fig. 3. This leads to a deterioration of transient stability with the increase of $K_p$.

In summary, the reduction of $K_i$ and increase of $K_p$ have a positive effect on the existence of $\omega_{c,cr}$ but may deteriorate the transient stability of the converter. A proper $K_i$ and $K_p$ should be selected in advance according to the SCR or other parameters of the system.

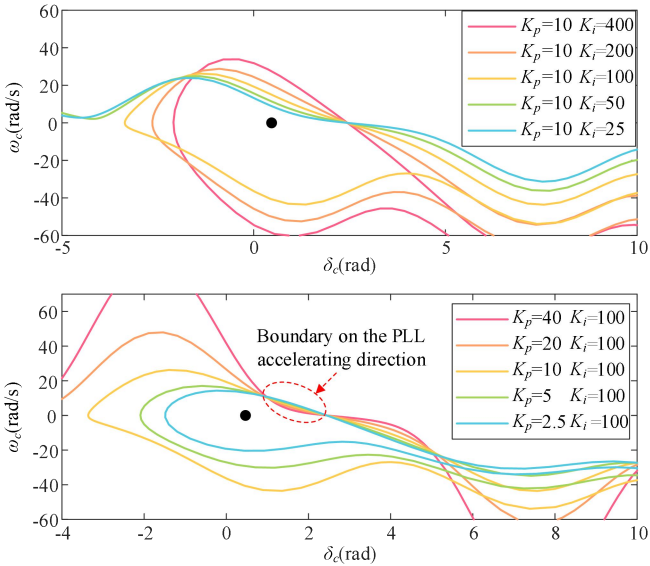

Fig. 3. SDs of GFL converter under different PLL parameters.

From (3), the current $I_c \angle \varphi_I$ also has a significant influence on transient stability. When a converter exits the low-voltage ride-through (LVRT) process after fault clearance, the q-axis current is generally set to be 0, which means $\varphi_I=0$ and $I_c=I_{cd}$. Using $I_{dref0}$ to represent the pre-fault d-axis current, SDs under different $I_{cd}$ are shown in Fig. 4. With the increase of $I_{cd}$, the SD shrinks both in $\delta_c$ dimension and $\omega_c$ dimension. This indicates that properly reducing the initial d-axis recovery current after LVRT exit can improve transient stability characteristics. However, the frequency problem may be prominent due to the reduction of active power output.

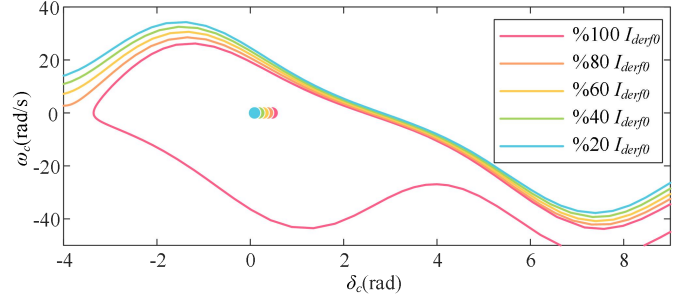

Fig. 4. SDs of the GFL converter under different $I_{cd}$.

*B. PLL-reset Control*

From the above analysis, adjusting control parameters cannot completely eliminate LOS and may lead to other stability issues. Unlike synchronous generators, whose phase angle is determined by the rotor's position, the phase angle of the GFL converter is determined by the phase-locked loop (PLL), serving merely as a control signal and lacking a real-world physical counterpart. When the state ($\delta_c$, $\omega_c$) lies outside the SD, the output of the PLL ($\omega_c$ or $\delta_c$) can be adjusted through specific control logic to bring the state back within the SD, thereby preventing instability.

From the analysis of SD in Section II.A, two PLL-reset control methods can be proposed:

1) **ω reset**: When the SCR is not very low, like the cases in the first and second figures of Fig. 2, $\omega_c$ is set to zero, or a specific value $\omega_{c,cr}$, and $\delta_c$ remains unchanged.

2) **ω&δ reset**: When the SCR is very low, like the case in the third figure of Fig. 2, both $\omega_c$ and $\delta_c$ are set to zero.

The control block diagram of the PLL-reset control is shown in Fig. 5. The sections shown in black lines are the traditional PLL. The sections in blue apply to both aforementioned PLL-reset methods, and the sections in purple are specific to the ω&δ reset method. The reset signal is a pulse signal, and when the PLL is reset, a positive edge will be generated. The positive edge will first reset the integrator output to the value $\omega_{c,cr} - U_{cq}K_p$, which makes $\omega_c$ reset to the specific value $\omega_{c,cr}$. If the ω&δ reset method is adopted, the output of the hold block will be held to the input value at the positive edge time. This means that $\delta_c$ will always subtract the phase angle at the reset time, which makes $\delta_c$ reset to zero.

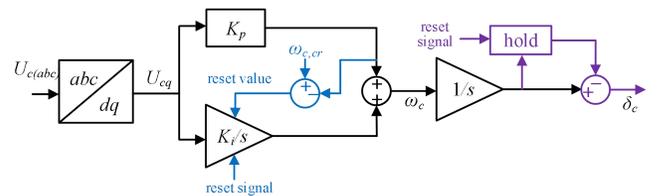

Fig. 5. Control block diagram of the PLL-reset control.

It is worth noting that the converter generally cannot get the phase angle of other devices. When the converter connects to a large power system, the phase angle of the system does not fluctuate significantly during the transient process. This is the prerequisite of the $\omega\&\delta$ reset. When the SG or other grid-forming devices are insufficient, the phase angle of the system also fluctuates, and resetting $\delta_c$ may lead to a large deviation between the PLL phase and the system phase and deteriorate the transient stability. In this scenario, only $\omega$ reset is appropriate.

With the PLL-reset control logic in Fig. 5, generating the reset signal is essential. A simple but effective method is to implement the stability judgment when the converter exits LVRT control. Though the TRM can give the precise SD of the converter, there is no formula method or criterion in the TRM to detect whether the state ($\delta_c$, $\omega_c$) is located in or out of the SD. This makes the stability judgment process difficult to embed in a control logic and, therefore, unsuitable for PLL-reset control. In Lyapunov-based methods, subscribe the current state ($\delta_c$, $\omega_c$) to the Lyapunov function $V(\delta_c, \omega_c)$, and if the value is larger than the critical value $V_{cr}$, the state is regarded as unstable. Only algebraic operation is involved in this judgment process, making it highly suitable for integration with control logic.

The Lyapunov-based PLL-reset control process is shown in Fig. 6. The Lyapunov method is a trigger for the reset control. The Lyapunov function $V(\delta_c, \omega_c)$ is established in advance. When the fault is cleared and the converter exits LVRT, the current state ($\delta_c$, $\omega_c$) is subscribed to $V(\delta_c, \omega_c)$ for stability judgment. If the judgment result is unstable, which means $V(\delta_c, \omega_c) > V_{cr}$, a positive edge is generated in the reset signal of Fig. 5, and the reset control is activated.

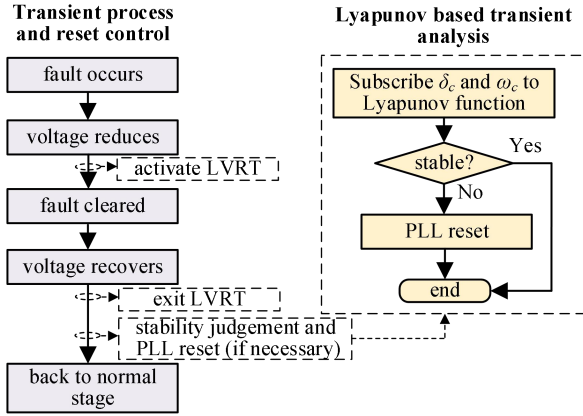

Fig. 6. Lyapunov-based PLL-reset control process.

## IV. Lyapunov-Based Transient Analysis Assessment

### A. The negative damping problem of the traditional Lyapunov method

From (3), the dynamic equations of the GFL converter have the same form as the swing equation of SGs. However, the damping coefficient varies with the phase angle $\delta_c$, and when $\delta_c < \theta_1$, the damping coefficient is negative. The traditional Lyapunov function is valid when the damping coefficient is positive, or the damping energy consumption/accumulation should be included. The traditional Lyapunov function $V_{tr}$ for the GFL converter with damping energy included can be expressed by:

$$V_{tr} = \frac{1}{2}\omega_c^2 - P_{m,c}(\delta_c - \delta_{c,se}) + \int_{\delta_{c,se}}^{\delta_c} D'_c \omega_c d\delta_c \\ - P_{e,c}\left(\cos(\delta_c - \theta_1) - \cos(\delta_{c,se} - \theta_1)\right) \quad (5)$$

Where, $\delta_{c,se}$ is the phase angle at the stable equilibrium point (SEP). It can be seen that the first, second, and last terms of the Lyapunov function are only related to the current state and SEP, but the value of the third damping term is also highly related to the trajectory during the transient process. This makes it difficult to determine the critical energy value from the traditional closest UEP method or the controlling UEP method [13].

### B. The approximation-based Lyapunov method

Two approximation methods can be used to calculate the path-dependent term in (5) (damping term), which are ray approximation and trapezoid approximation. The procedure for ABLM is shown below.

**Step 1:** Obtain the analytical expression of the path-dependent terms. The process of ray approximation is as follows. When only the original and final states are considered, the variables in the damping term are approximated as:

$$\delta_c = \delta_{c,se} + \lambda\Delta\delta_c = \delta_{c,se} + \lambda(\delta_c - \delta_{c,se}), \\ \omega_c = \lambda\Delta\omega_c = \lambda(\omega_c - 0) = \lambda\omega_c \quad (6)$$

Where, $\lambda$ is a substituted variable within the range [0, 1]. $\Delta\delta_c$ and $\Delta\omega_c$ are the distance from the current state to SEP. From (6), the number of variables in the damping term is reduced to one, and the definite integral can be analytically solved. By the ray approximation, the damping term $E_{damp}$ can be rewritten as:

$$E_{d,r} = \int_0^1 D_c \cos(\delta_{c,se} + \lambda\Delta\delta_c - \theta_1)\lambda\Delta\omega_c\Delta\delta_c d\lambda \\ = D_c \omega_c \sin(\delta_c - \theta_1) \\ + D_c\left[\cos(\delta_c - \theta_1) - \cos(\delta_{c,se} - \theta_1)\right]\frac{\omega_c}{\delta_c - \delta_{c,se}} \quad (7)$$

Trapezoid approximation is based on the closed Newton-Cotes formula. It estimates the integration by transferring the original integration area of each step to a trapezoid and calculating the total area of the trapezoids as the approximation to the integration. The trapezoid approximation-based damping term $E_{d,tr}$ can be rewritten as:

$$E_{d,tr} = \frac{1}{2} * \sum_{i=1}^{N}\left(D'_{c,i+1}(\omega_{c,i+1} - \omega_0)^2 + D'_{c,i}(\omega_{c,i} - \omega_0)^2\right)\Delta t \\ D'_{c,i} = D_c \cos(\delta_{c,i} - \theta_1) \quad (8)$$

Where, $N$ is the number of the steps of the approximation. $\Delta t$ is the time interval of each step, and $\Delta\delta_{c,i}$ is the phase angle interval of step $i$. Though the multi-step Trapezoid approximation is given in (8), it is worth noting that only the single-step trapezoid approximation (i.e. $N=1$) form of $E_{d,tr}$ is trajectory-free, which means $E_{d,tr}$ is only determined by the start and end of the trajectory.

These two approximation methods are based on the assumption that the trajectory from SEP to UEP is the segment between the two points. However, due to the mutation of PLL output, the actual trajectory is significantly different from the segment between SEP and UEP, which may increase the error.

*C. The Zubov-based method*

The Zubov-based method is another way to generate a Lyapunov function $V_{zu}$, and it has been proved that the condition $V_{zu}=1$ is a family curve of $\partial A$. $V_{zu}$ can be expressed as:

$$\dot{V}_{zu} = \boldsymbol{F} \frac{\partial V_{zu}}{\partial \boldsymbol{x}} = -\phi(1-V_{zu}) \quad (9)$$

Where, $\boldsymbol{F}$ is the dynamic equations in (3). $\boldsymbol{x}$ is the vector of the variables $[\delta_c, \omega_c]$. $\phi$ is an arbitrary positive defined function of $\delta_c$ and $\omega_c$. $V_{zu}$ can be obtained only when the dynamic equations are polynomial equations, and Taylor expansion can be used to get an approximation of $V_{zu}$, which is represented by $V_{zu}^{(M)}$; the superscript $(M)$ means that the original $V_{zu}$ is truncated at degree $M$. Unlike $V_{zu}$, $V_{zu}^{(M)}=1$ is no longer the accurate estimation of $\partial A$, but a critical energy value can be found to guarantee the conservative of the estimation and minimize the errors. More details can be found in [12]. The procedure of the Zubov-based method is shown as follows.

**Step 1:** obtain the polynomial form of dynamic equations by Taylor Expansion. Use $f_n$ to represent the $n^{th}$ equation of $\boldsymbol{F}$, and $f_n^{(M_T)}$ represents the truncated $f_n$ at degree $M_T$ after Taylor Expansion. $f_n^{(M_T)}$ can be expressed by:

$$f_n^{(M_T)} = \sum_{k=1}^{2} b_{nk} x_k + f_n'^{(M_T)} \quad (10)$$

Where, $b_{n1}$ and $b_{n2}$ are the coefficients of linear terms. $x_1$ and $x_2$ are $\delta_c$ and $\omega_c$ in this paper. $f_n'^{(M_T)}$ is the rest terms of $f_n^{(M_T)}$.

**Step 2:** from (9), $V_{zu}^{(M)}$, which is truncated at degree $m$, can be calculated by the following recursion process:

$$\sum_{n=1}^{2} \frac{\partial V_{zu,2}^{(M)}}{\partial x_n} \left( \sum_{k=1}^{2} b_{nk} x_k \right) = -\phi \quad (11)$$

$$\sum_{n=1}^{2} \frac{\partial V_{zu,m}^{(M)}}{\partial x_n} \left( \sum_{k=1}^{2} b_{nk} x_k \right) = R_m(x) \quad m=3,4,5,...,M \quad (12)$$

Where, $V_{zu,m}^{(M)}$ is the $m$-degree homogeneous multinomial in $V_{zu}^{(M)}$. $R_m(x)$ is a $m$-degree homogeneous multinomial obtained from (9) and the Taylor expansion of $F$.

**Step 3**: obtain the set of $dV_{zu}^{(M)}/dt = 0$, and find the minimum $V_{zu}^{(M)}$ value $c_{zu}$ in this set. The set $V_{zu}^{(M)} = c_{zu}$ is the estimated $\partial A$.

**Step 4**: substitute the state $(\delta_c, \omega_c)$ to the Lyapunov function $V_{zu}^{(M)}$, and if $V_{zu}^{(M)}(\delta_c, \omega_c) \leq c_{zu}$, $(\delta_c, \omega_c)$ is in the stability domain $A$.

The Zubov method can provide a fixed procedure for the construction of Lyapunov functions. There is no specific requirement for the form of the dynamic equations, which makes the other system dynamics or control strategies of the GFL converter easy to include. The main problem is that for high-order systems, the number of variables in the Lyapunov function increases significantly, which also significantly increases the time required for Lyapunov function construction.

*D. The analytical trajectory reversing method*

The basic concept of ATRM is to expand an estimated $A$ from the trajectory-reversing perspective. Suppose the original estimated $\partial A$ is:

$$V_{atr}(\boldsymbol{x}) = k_0 \quad (13)$$

Where, $V_{atr}$ is the Lyapunov function of ATRM, and variables $x=(\delta_c, \omega_c)$. $k_0$ is a constant. The expansion of the boundary in (13) can be expressed as:

$$V_{atr}(\boldsymbol{x}) + \left(\frac{\partial V_{atr}}{\partial \boldsymbol{x}}\right)^T \boldsymbol{F} dt = k_0 \quad (14)$$

Where, $\boldsymbol{F}$ is the dynamic equations in (3), and $dt$ is the time interval of the trajectory reversing.

The procedure of calculating $V_{atr}$ and boundary expansion procedure is given in [14], and $V_{atr}$ can be expressed by:

$$V_{atr}(\boldsymbol{x}) = V_{atr}(\boldsymbol{x}, \boldsymbol{p}) = \sum_{i=1}^{M} \boldsymbol{v}_i^T(\boldsymbol{x}) \boldsymbol{p}_i \quad (15)$$

Where, $\boldsymbol{p}$ is a set of coefficients of the polynomial $V_{atr}$. $\boldsymbol{v}_i = [\delta_c^i, \delta_c^{i-1}\omega_c, ..., \delta_c\omega_c^{i-1}, \omega_c^i]$, and $\boldsymbol{p}_i$ is the vector of coefficients of homogeneous multinomials in $V_{atr}$. $V_{atr}(\boldsymbol{x}, \boldsymbol{p})$ also satisfies:

$$\left(\frac{\partial V_{atr}}{\partial \boldsymbol{p}}\right)^T \frac{d\boldsymbol{p}}{dt} = \left(\frac{\partial V_{atr}}{\partial \boldsymbol{x}}\right)^T \boldsymbol{F} \quad (16)$$

The procedure of ATRM is shown as follows:

**Step 1:** solve the $M_i$-degree polynomial Lyapunov function $V_{atr}$. Firstly, use (16) to recursively solve the analytical expression of $\boldsymbol{p}_i$ with time $t$ as the independent variable. Secondly, find the minimum $t$ satisfying $\dot{V}_{atr}(\boldsymbol{x}, \boldsymbol{p}_i(t)) = 0$ and $V_{atr}(\boldsymbol{x}, \boldsymbol{p}_i(t)) = k_0$, which is represented by $t_s$. The region $V_{atr}(\boldsymbol{x}, \boldsymbol{p}_i(t_s)) \leq k_0$ is the estimation of SD $A$.

**Step 2:** assume $M_{i+1} > M_i$ as the maximum degree of $V_{atr}$. The dimension of $\boldsymbol{p}$ also increases, and the original value of $\boldsymbol{p}$ at $t=0$ can be calculated by

$$\boldsymbol{p}_{i+1}(0) = \left[\boldsymbol{p}_i(t_s - \varepsilon_0), \overbrace{0, ..., 0}^{M_{i+1}}\right] \quad (17)$$

Where, $\boldsymbol{p}_i$ is obtained in Step 1. $\varepsilon_0$ is a small time interval. With the new $\boldsymbol{p}_{i+1}(0)$, go back to Step 1 and calculate the new $V_{atr}$ for (16), until the estimation accuracy satisfies the requirements.

Like the Zubov method, the ATRM provides another fixed procedure for the construction of the Lyapunov function, and it is also feasible when other dynamics of the power system or strategies of the GFL converter are considered. The estimated boundary approaches the real boundary $\partial A$ after each expansion, but in the Zubov method, an increase of the degree $M$ of the Lyapunov function $V_{zu}^{(M)}$ cannot guarantee a more accurate estimation. This means ATRM has better convergence characteristics than the Zubov method. However, some of the procedures in ATRM are much more complicated than those in the Zubov method, like obtaining analytical solutions of differential equations for $p$ or finding $t_s$.

*E. Comparison of the three proposed methods*

A comparison of the methods is given in Table I. TRM is the most accurate method but mainly for second-order systems, which means including some control strategies or dynamics may be feasible, and the system is still a second-order system. Another problem for TRM is that judging the stability of a state is slow. The problem with the ABLM is that the radical error may be introduced, and the dynamic equations should be in the form of a swing equation. The characteristics of the Zubov method and ATRM are similar, but the calculation procedure of ATRM is more complicated. Compared to the ABLM, these two methods only require that the dynamic equations be Taylor-expanded, which results in better expansibility.

TABLE I.
COMPARISON OF THE TRANSIENT ANALYSIS METHODS

| Method | Characteristic | Description |
|---|---|---|
| TRM | Accuracy | Almost no error. |
| | Speed | Fast to estimate the SD, but slow to judge the stability of one state. |
| | Expansibility | Mainly for second-order systems. |
| ABLM | Accuracy | Both radical and conservative errors exist. |
| | Speed | Fast to estimate the SD or critical energy value, and fast to judge. |
| | Expansibility | Hard to include other dynamics or control strategies. |
| Zubov method | Accuracy | Conservative error exists. |
| | Speed | Slow to estimate the SD, but fast to judge. |
| | Expansibility | Applicable when other dynamics or control strategies are included but not suitable for high-order systems. |
| ATRM | Accuracy | Conservative error exists. |
| | Speed | Slow to estimate the SD, but fast to judge. |
| | Expansibility | Applicable when other dynamics or control strategies are included but not suitable for high-order systems. |

## V. CASE STUDY

*A. Comparison of different Lyapunov-based methods*

In this subsection, the ordinary differential equation (ODE) based simulation model of the system in Fig. 1 is established in MATLAB, and the transient analysis method is also conducted. The parameters of the system are given in the Appendix. The result of TRM is used as the real boundary to analyze the accuracy of other methods. For convenience of demonstration, the original variable $\delta_c$ is replaced by $\delta_c-\delta_{c,se}$ in Section V, which makes (0,0) the SEP. The moment before CCT and estimated CCT are represented as $t_{CCT-}$ and $t_{eCCT-}$.

Firstly, ray approximation and trapezoid approximation-based ABLM are used for the transient analysis. The trapezoid approximation is single-step, which means $N=1$ in (8). The estimated stability boundary, the real stability boundary, and the trajectory when a fault is cleared at $t_{CCT-}$ are shown in Fig. 7. It can be seen that the ray approximation-based method has a relatively smaller error. However, the conservative property of the estimation cannot be guaranteed by either method. From the fault trajectory, the exit point is on the stability boundary but within the estimated boundary of the two methods, and it is foreseeable that the estimated CCT will be radical.

The estimation of CCT under different fault resistances by two approximation methods is shown in Table II. Estimations of both two methods are radical (negative error). The error of the ray approximation method is around -5%, and the error of the trapezoid approximation method is around -65%.

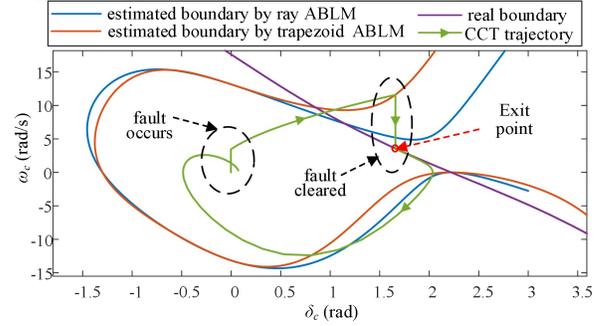

Fig. 7. Transient analysis results of ABLM.

TABLE II.
CCT ESTIMATION UNDER DIFFERENT FAULT RESISTANCES

| $R_f$ | 3Ω | 1Ω | 0.5Ω | 0.1Ω |
|---|---|---|---|---|
| Real CCT | 0.2319s | 0.2235s | 0.2218s | 0.2203s |
| Ray | 0.2432s | 0.2354s | 0.2338s | 0.2325s |
| Error | **-4.873%** | **-5.324%** | **-5.410%** | **-5.538%** |
| Trapezoid | 0.3742s | 0.3703s | 0.3692s | 0.3688s |
| Error | **-61.363%** | **-65.638%** | **-66.456%** | **-67.408%** |

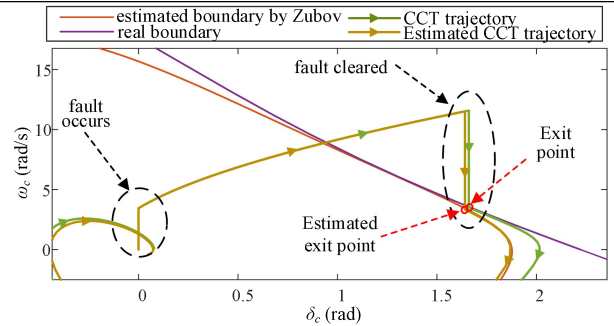

Fig. 8. Transient analysis results of the Zubov method.

TABLE III.
CCT ESTIMATION UNDER DIFFERENT FAULT RESISTANCES

| $R_f$ | 3Ω | 1Ω | 0.5Ω | 0.1Ω |
|---|---|---|---|---|
| Real CCT | 0.2319s | 0.2235s | 0.2218s | 0.2203s |
| Estimated CCT | 0.2303s | 0.2221s | 0.2200s | 0.2186s |
| Error | **0.690%** | **0.630%** | **0.811%** | **0.771%** |

With the Zubov method, the estimated stability boundary, the real stability boundary, and the trajectories when the fault is cleared at $t_{CCT-}$ and $t_{eCCT-}$ are shown in Fig. 8. The Lyapunov function $V_{zu}$ is truncated at degree $M=16$. The results show that the estimated boundary is entirely within the actual boundary, which demonstrates the conservativeness of the Zubov method. The actual CCT and the estimated CCT are 0.2235 seconds and 0.2221 seconds, respectively, with a small error of 0.630%. To further evaluate the accuracy of the Zubov method, CCT is estimated under different fault resistances $(R_f)$, and the results are presented in Table III. From Table III, the estimation error for all cases is under 1%, and no radical errors are observed, reflecting the good performance of the Zubov method in the case study.

The estimated stability boundary from ATRM, the real stability boundary, and the trajectories when the fault is cleared at $t_{CCT}$- and $t_{eCCT}$- are shown in Fig. 9. The original boundary is $\delta_c^2 + 0.01\omega_c^2 = 0.8$, and after one expansion by ATRM, the estimated boundary greatly approaches the real boundary. However, when conducting the second expansion, the differential equation (16) becomes too complicated to obtain a solution. The results of CCT estimation under different fault resistances $R_f$ are shown in Table IV. The estimation errors in different cases are all under 9%, and no radical error is observed.

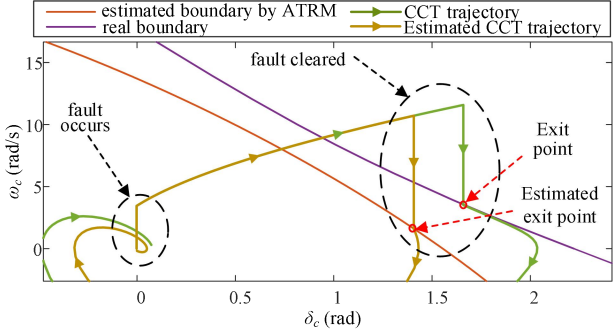

Fig. 9. Transient analysis results of ATRM.

TABLE IV.
CCT ESTIMATION UNDER DIFFERENT FAULT RESISTANCES

| $R_f$ | 3Ω | 1Ω | 0.5Ω | 0.1Ω |
|---|---|---|---|---|
| Real CCT | 0.2319s | 0.2235s | 0.2218s | 0.2203s |
| Estimated CCT | 0.2132s | 0.2044s | 0.2024s | 0.2007s |
| Error | **8.103%** | **8.546%** | **8.747%** | **8.897%** |

The computational time of different methods is shown in Table V. Results show that the Zubov method and ATRM are time-consuming, and TRM and ABLM are much faster, which corresponds with the theoretical analysis.

TABLE V. TIME CONSUMPTION OF DIFFERENT METHODS

| | TRM | ABLM (ray) | ABLM (trapezoid) | Zubov ($M$=16) | ATRM (one expansion) |
|---|---|---|---|---|---|
| Time | 1.14s | 1.34s | 1.12s | 6.83s | 10.78s |

*B. Performance of the PLL-reset control in a single-machine system*

In this subsection, an electromagnetic simulation model of the system in Fig. 1 is developed in the PSCAD/EMTDC software program. Both PLL-reset methods in Section III.B are tested in the simulation. The LVRT control method and parameters are detailed in the Table VI, Appendix. From the comparison in Section V.A, the Zubov method has the highest accuracy and is chosen as the stability judgment method in the PLL-reset control.

Firstly, the converter side impedance $Z_c$ is 0+j0.36p.u., and the SCR is 2.1195. From Fig. 2, there exists an $\omega_{c,cr}$ to guarantee the stability of the converter regardless of $\delta_c$, which is set to be -2π in this simulation. Simulation results with and without the $\omega$ reset control are compared in Fig. 10, containing the PCC voltage amplitude $U_c$, $\delta_c$, and PLL output frequency. The fault occurs at time 0.2s. The blue and orange lines represent the simulation results for a fault duration of 0.27 seconds, while the yellow lines represent the results for a fault duration of 1 second.

In the first test, cases with a fault duration of 0.27 seconds are analyzed. 0.095s after the fault clearance, the PCC voltage recovers to 0.9 p.u., which makes the converter exit LVRT control and activates the stability judgment in Fig. 6. Substituting the state (2.5036, 10.7548) into the Zubov-based Lyapunov function, $V_{zu}^{(16)}$ (2.5036, 10.7548)= 199.845, which is greatly larger than the critic value $c_{zu}$ =0.558, and the PLL-reset is activated. From Fig. 10 (c), at the reset activation time, the frequency is reset to 49Hz, which corresponds to parameter $\omega_{c,cr}$=-2π. From the comparison in Fig. 10, when the PLL-reset control is not used, the PLL output frequency keeps increasing until it reaches the upper bound, which enlarges the phase angle $\delta_c$, and leads to loss of synchronization. By resetting $\omega_c$ to a negative value (i.e., resetting frequency to be less than 50Hz), $\delta_c$ immediately reduces, which avoids LOS. From the yellow lines in Fig. 10, when the fault duration is increased to 1s, transient stability can still be achieved through frequency reset. This further verifies the effectiveness of the reset control.

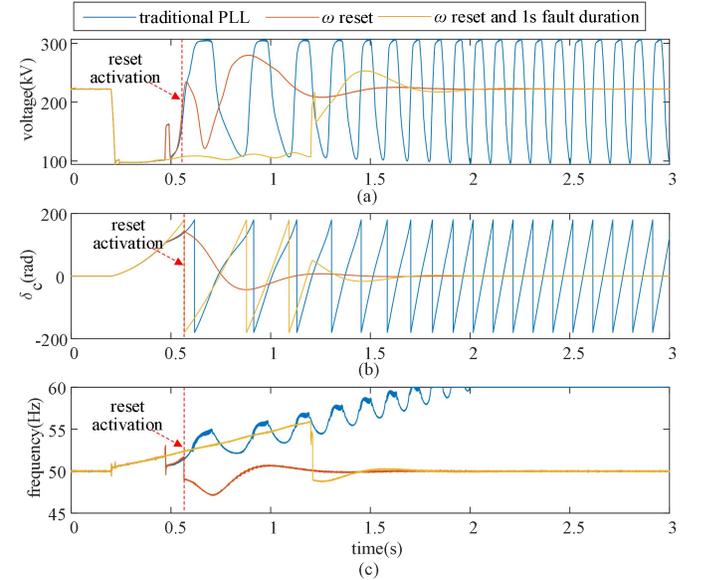

Fig. 10. Comparison of the traditional PLL control and the $\omega$ reset control.

In the second test, the converter side impedance $Z_c$ is further increased to 0+j0.6p.u., leading to an extremely low SCR=1.4054. From Fig. 2, $\omega_{c,cr}$ guaranteeing the stability of the converter no longer exists. The $\omega$ reset and $\omega\&\delta$ reset control are compared, and in the $\omega$ reset control, $\omega_c$ is also reset to -2π like the last test, and in the $\omega\&\delta$ reset control, $\omega_c$ and $\delta_c$ are reset to 0. Simulations under the two reset methods are compared in Fig. 11, containing the PCC voltage amplitude $U_c$, $\delta_c$, and PLL output frequency. The fault occurs at time 0.2s. The blue and orange lines show the simulation results for a fault duration of 0.232s, while the yellow lines show the results for a fault duration of 1s.

The cases with a fault duration of 0.232s, are firstly analyzed. 0.003s after the fault clearance, The PCC voltage recovers to 0.9 p.u., and the converter exits LVRT control. The stability judgment in Fig. 6 is also activated. Substituting the state (3.1000, 24.7785) into the Zubov-based Lyapunov function, $V_{zu}^{(16)}$ (3.1000, 24.7785) =5.422×10$^7$, which is significantly larger than the critical value $c_{zu}$ =0.334, and the PLL-reset is activated. In Fig. 11(c), when the reset is activated, the frequency is reset to 49 Hz by the $\omega$ reset control and 50 Hz by the $\omega\&\delta$ reset control, which corresponds to their reset values of -2π and 0. Under $\omega$ reset control, the PLL output frequency continues to increase until it reaches the upper bound, which in turn enlarges the phase angle $\delta_c$, leading to LOS. However, under $\omega\&\delta$ reset control, both $\omega_c$ and $\delta_c$ are reset to 0. The frequency quickly mutates to 52 Hz and finally returns to the SEP, achieving synchronization. From the yellow lines in Fig. 11, the fault duration is increased to 1s, but the transient stability can still be achieved through $\omega\&\delta$ reset. This further verifies the effectiveness of the reset control.

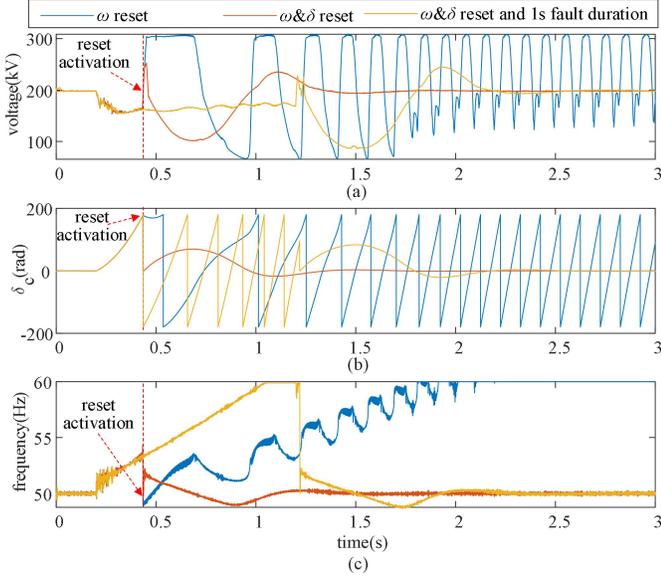

Fig. 11. Comparison of the $\omega$ reset control and the $\omega\&\delta$ reset control.

*C. Performance of the PLL-Reset control in a multi-machine system*

The two reset control methods are also tested in a modified IEEE 9-bus system using the PSCAD/EMTDC software. The topology is shown in Fig. 12. The basic line and load parameters are in [24]. Based on the system in [24], the SG in bus 2 is replaced by a GFL converter, and an additional line with an impedance of 0.03+j0.3 p.u. is added between bus 2 and bus 7 to weaken the system's strength. The loads are impedance loads, and the other parameters of the dynamic model are shown in Table VII in the Appendix.

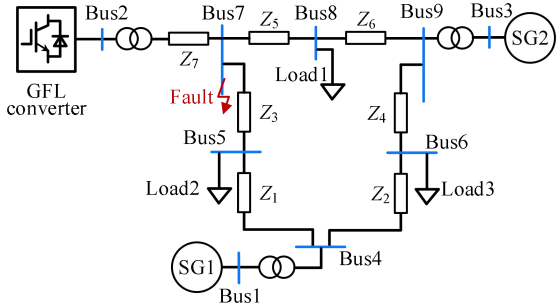

Fig. 12. Topology of the modified IEEE 9 bus system

Simulation results under a three-phase short-circuit fault at bus 7 with a 0.1s duration are shown in Fig. 13. Two resets are activated 0.02s after the fault clearance, and $\omega_c$ is reset to 0 in $\omega$ reset control. Without PLL-reset control, the LOS occurs after fault clearance, and both reset controls can achieve synchronization after fault clearance. Comparing the post-fault voltage curve, voltage fluctuation is larger under $\omega$ reset control. This is because there is still a large deviation between $\delta_c$ and the SEP, leading to a large active and reactive power deviation compared to the rated value. In contrast, the $\omega\&\delta$ reset has better performance in alleviating power and voltage fluctuation. From the $\delta_c$ curve, the final $\delta_c$ value is close to the pre-fault value, indicating that the phase angles of the SGs do not undergo significant changes during transient processes. So, resetting the $\delta_c$ to zero is feasible. When the capacity of the SGs decreases, the phase angle of the SG also fluctuates significantly. If $\delta_c$ is still reset to zero, a significant power angle difference may exist between the converter and the SG. Therefore, when connecting to a large system with a stable phase angle, $\omega\&\delta$ reset control is recommended to minimize post-fault fluctuations. When the phase angle of the system also fluctuates prominently in the transient process, $\omega$ reset control is recommended.

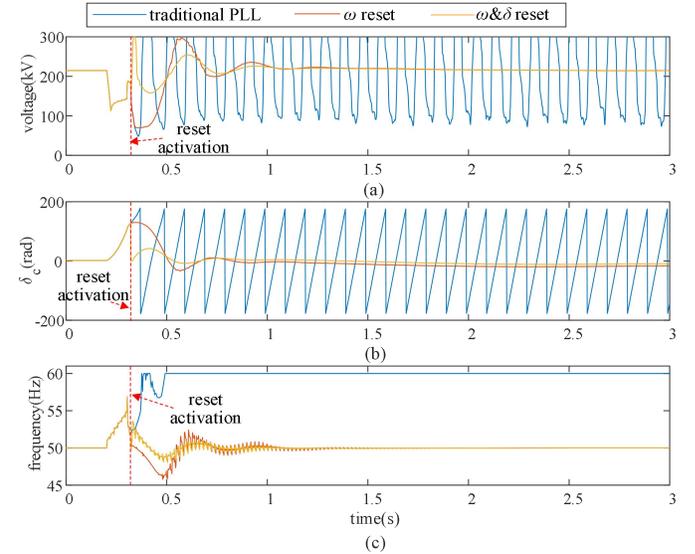

Fig. 13. Comparison of the basic PLL control, $\omega$ reset control, and $\omega\&\delta$ reset control in a multi-machine system.

## VI. CONCLUSION

To prevent the loss of operation (LOS) of GFL converters under large disturbances, a PLL-reset control is proposed in this paper. The following conclusions can be drawn from the theoretical analysis and case studies:

1) The accurate SD of the GFL converter under constant-current control is given by the TRM. Under certain control parameters, except for the very low SCR scenario, a frequency (i.e., angular velocity) value can be found for a constant-current controlled GFL converter so that no matter how the phase angle changes, as long as the frequency is lower than the specific value, the converter is in the stable region.

2) Based on the first conclusion, two kinds of PLL-reset control logic are proposed. By resetting the PLL frequency or both frequency and phase angle, the GFL converter can avoid LOS. The effectiveness of the control is verified in case studies. Results show that $\omega$ reset can guarantee stability for the first and second SD form cases, and $\omega\&\delta$ reset can guarantee all three SD form cases.

3) Three Lyapunov function construction methods are

proposed as the triggering condition of the reset control. Among the three methods, the Zubov and ITRM methods have higher accuracy, and conservative estimation is guaranteed. The ABLM method may have optimistic errors, so it is not recommended.

## APPENDIX

TABLE VI.
PARAMETERS OF THE SINGLE MACHINE TEST SYSTEM

| Parameter | Explanation | Value |
|---|---|---|
| $S_b, U_b, \omega_0$ | Base value of power, voltage, and angular velocity. | 100MVA, 230kV, 100π rad/s |
| $R_c, X_c$ | Resistance and reactance of $Z_c$. | 0p.u., 0.36p.u. |
| $R_g, X_g$ | Resistance and reactance of $Z_g$. | 0p.u., 0.12p.u. |
| $R_l, X_l$ | Resistance and reactance of $Z_l$. | 0.5p.u., 5p.u. |
| $R_f$ | Fault resistance. | 1Ω |
| $U_g$ | Grid voltage. | 1.1p.u. |
| $K_i, K_p$ | Integral/proportional coefficient of PLL. | 100, 10 |
| $I_c, \varphi_I$ | Converter current and its phase angle | 1p.u., 0 rad |

TABLE VII.
PARAMETERS OF THE MODIFIED IEEE 9 BUS TEST SYSTEM

| Parameter | Explanation | Value |
|---|---|---|
| $S_b, U_b, \omega_0$ | Base value of power, voltage, and angular velocity. | 100MVA, 230kV, 100π rad/s |
| $Z_7$ | Resistance and reactance of the additional impedance between bus 2 and bus 7. | 0.03+j0.3p.u. |
| $R_f$ | Fault resistance. | 1Ω |
| $S_c$ | Capacity of GFL resource | 200MVA |
| $K_i, K_p$ | Integral/proportional coefficient of PLL. | 500, 20 |
| $I_c, \varphi_I$ | Reference converter current and its phase angle ($P_c$=163MW, $Q_c$=20MVar) | 0.817p.u., 0.123rad |
| $S_g$ | Capacity of SGs | 300MVA |
| $J_g, K_g$ | Inertia and frequency droop of SGs | 4s, 20p.u. |

## REFERENCES

[1] Q. Hu, L. Fu, F. Ma and F. Ji, "Large Signal Synchronizing Instability of PLL-Based VSC Connected to Weak AC Grid," in *IEEE Transactions on Power Systems*, vol. 34, no. 4, pp. 3220-3229, July 2019.
[2] "Odessa Disturbance," Joint NERC and Texas RE Staff Report, Sep. 2021. [Online]. Available: https://www.nerc.com/pa/rrm/ea/Documents/Odessa_Disturbance_Report.pdf
[3] X. He, H. Geng, R. Li and B. C. Pal, "Transient Stability Analysis and Enhancement of Renewable Energy Conversion System During LVRT," in *IEEE Transactions on Sustainable Energy*, vol. 11, no. 3, pp. 1612-1623, July 2020.
[4] Z. Tian et al., "Hamilton-Based Stability Criterion and Attraction Region Estimation for Grid-Tied Inverters Under Large-Signal Disturbances," in *IEEE Journal of Emerging and Selected Topics in Power Electronics*, vol. 10, no. 1, pp. 413-423, Feb. 2022.
[5] Z. Tian et al., "Transient Synchronization Stability of an Islanded AC Microgrid Considering Interactions Between Grid-Forming and Grid-Following Converters," in *IEEE Journal of Emerging and Selected Topics in Power Electronics*, vol. 11, no. 4, pp. 4463-4476, Aug. 2023, doi: 10.1109/JESTPE.2023.3271418.
[6] Y. Ma, D. Zhu, Z. Zhang, X. Zou, J. Hu, and Y. Kang, "Modeling and Transient Stability Analysis for Type-3 Wind Turbines Using Singular Perturbation and Lyapunov Methods," in *IEEE Transactions on Industrial Electronics*, vol. 70, no. 8, pp. 8075-8086, Aug. 2023.
[7] X. Li, Z. Tian, X. Zha, P. Sun, Y. Hu and M. Huang, "An Iterative Equal Area Criterion for Transient Stability Analysis of Grid-Tied Converter Systems with Varying Damping," in *IEEE Transactions on Power Systems*, vol. 39, no. 1, pp. 1771-1784, Jan. 2024.
[8] C. He, X. He, H. Geng, H. Sun, and S. Xu, "Transient Stability of Low-Inertia Power Systems with Inverter-Based Generation," in *IEEE Transactions on Energy Conversion*, vol. 37, no. 4, pp. 2903-2912, Dec. 2022.
[9] Q. -H. Wu, Y. Lin, C. Hong, Y. Su, T. Wen, and Y. Liu, "Transient Stability Analysis of Large-scale Power Systems: A Survey," in *CSEE Journal of Power and Energy Systems*, vol. 9, no. 4, pp. 1284-1300, July 2023.
[10] Z. Li, J. Li, D. Gan, and Z. Wang, "Stability Analysis of PLL-Type Grid-Connected Converter Based on The Perturbation Method," in *IEEE Transactions on Power Delivery*, vol. 39, no. 2, pp. 1151-1161, April 2024
[11] L. Zheng, X. Liu, Y. Xu, W. Hu, and C. Liu, "Data-driven Estimation for a Region of Attraction for Transient Stability Using the Koopman Operator," in *CSEE Journal of Power and Energy Systems*, vol. 9, no. 4, pp. 1405-1413, July 2023.
[12] Y. Yu and K. Vongsuriya, "Nonlinear Power System Stability Study by Liapunov Function and Zubov's Method," in *IEEE Transactions on Power Apparatus and Systems*, vol. PAS-86, no. 12, pp. 1480-1485, Dec. 1967.
[13] Hsiao-Dong Chiang and Chia-Chia Chu, "Theoretical foundation of the BCU method for direct stability analysis of network-reduction power system. Models with small transfer conductances," in *IEEE Transactions on Circuits and Systems I: Fundamental Theory and Applications*, vol. 42, no. 5, pp. 252-265, May 1995.
[14] R. Genesio and A. Vicino, "New techniques for constructing asymptotic stability regions for nonlinear systems," in *IEEE Transactions on Circuits and Systems*, vol. 31, no. 6, pp. 574-581, June 1984.
[15] C. Wu, X. Xiong, M. G. Taul, and F. Blaabjerg, "On the Equilibrium Points in Three-Phase PLL Based on the d-axis Voltage Normalization," in *IEEE Transactions on Power Electronics*, vol. 36, no. 11, pp. 12146-12150, Nov. 2021.
[16] Z. Dai, G. Li, Y. Yang, G. Wang, B. Li, and Y. Wang, "A Complex-Type Voltage Normalization Method for Three-Phase Synchronous Reference Frame Phase-Locked Loop," in *IEEE Transactions on Power Electronics*, vol. 39, no. 5, pp. 5873-5882, May 2024.
[17] M. Zhang and L. Sun, "PLL and Additional Frequency Control Constituting an Adaptive Synchronization Mechanism for VSCs," in *IEEE Transactions on Power Systems*, vol. 35, no. 6, pp. 4920-4923, Nov. 2020.
[18] J. Liu, X. Du, and Y. Zhao, "A Novel Control Strategy for Enhancing System Stability in Weak Grids by Mitigating Additional Disturbance Components from PLL," in *IEEE Transactions on Sustainable Energy*, vol. 16, no. 2, pp. 1114-1124, April 2025.
[19] Z. A. Alexakis, A. T. Alexandridis, P. C. Papageorgiou, and G. C. Konstantopoulos, "Design of a Bandwidth Limiting PLL for Grid-Tied Inverters With Guaranteed Stability," in *IEEE Transactions on Sustainable Energy*, vol. 16, no. 2, pp. 774-784, April 2025.
[20] H. Wu and X. Wang, "Design-Oriented Transient Stability Analysis of PLL-Synchronized Voltage-Source Converters," in *IEEE Transactions on Power Electronics*, vol. 35, no. 4, pp. 3573-3589, April 2020.
[21] Y. Ma, D. Zhu, J. Hu, R. Liu, X. Zou, and Y. Kang, "Optimized Design of Demagnetization Control for DFIG-Based Wind Turbines to Enhance Transient Stability During Weak Grid Faults," in *IEEE Transactions on Power Electronics*, vol. 40, no. 1, pp. 76-81, Jan. 2025.
[22] Y. Gu and T. C. Green, "Power System Stability With a High Penetration of Inverter-Based Resources," in *Proceedings of the IEEE*, vol. 111, no. 7, pp. 832-853, July 2023.
[23] L. Liu et al., "Preset power based droop control for improving primary frequency regulation of inverters under large disturbances," in *IEEE Transactions on Power Electronics*, vol. 40, no. 7, pp. 9153-9166, July 2025.
[24] Illinois Center for a Smarter Electric Grid. (2013). [Online]. Available FTP:http://publish.illinois.edu/smartergrid/